\documentclass[12pt]{article}
\usepackage[sort&compress,square,comma,numbers]{natbib}
\usepackage{braket,amsmath,amssymb,graphicx,slashed,mathtools}
\usepackage{accents,dsfont}
\usepackage[toc,page]{appendix}
\usepackage[export]{adjustbox}
\pdfoutput=1

\usepackage[
	colorlinks=true,
	citecolor=black,
	linkcolor=black,
	urlcolor=blue,
	hypertexnames=false]{hyperref}
	
\newcommand{\arXiv}[2]{\href{http://arxiv.org/pdf/#1}{{\tt #2/#1}}}
\newcommand{\arXivold}[1]{\href{http://arxiv.org/pdf/#1}{{\tt #1}}}

\newcommand{\beq}{\begin{equation}}
\newcommand{\eeq}{\end{equation}}

\newcommand{\V}[1]{\mathbf{#1}}

\begin{document}
\begin{center} 
{\huge \bf The Constructive Method for \\ \vspace*{0.25cm}} 
{\huge \bf Massive Particles in QED}
\end{center}

\begin{center} 
{\bf Hsing-Yi Lai,$^{a}$} {\bf Da Liu,$^{a,b}$} and {\bf John Terning$^a$}  \\
\end{center}
\vskip 8pt
\begin{center} 
{\it $^a$Center for Quantum Mathematics and Physics (QMAP),\\Department of Physics, University of California, Davis, CA 95616 \\
\it $^b$PITT PACC, University of Pittsburgh, Pittsburgh, PA, USA\\} 
\end{center}

\vspace*{0.1cm}
\begin{center} 
{\tt  \href{mailto:hylai@ucdavis.edu}{hylai@ucdavis.edu}\,,
\href{mailto:liudaphysics@gmail.com}{liudaphysics@gmail.com}\,,
 \href{mailto:jterning@gmail.com}{jterning@gmail.com}}

\end{center}
\centerline{\large\bf Abstract}
\begin{quote}
The constructive method of determining amplitudes from on-shell pole structure has been shown to be promising for calculating amplitudes in a more efficient way. However, challenges have been encountered when a massless internal photon is involved in the gluing of three-point amplitudes with massive external particles. In this paper, we use the original on-shell method, old-fashioned perturbation theory,  to shed light on the constructive method, and show that one can derive the Feynman amplitude by correctly identifying the residue even when there is an internal photon involved.
\end{quote}


\section{Introduction}
The BCFW recursion relations~\cite{Britto:2004ap,Britto:2005fq} provided an efficient way to calculate  tree-amplitudes of gluons from on-shell poles using complexified momenta. This has motivated much effort towards a better understanding of the scattering amplitudes~\cite{Arkani-Hamed:2008owk, Arkani-Hamed:2009ljj, Elvang:2013cua, Arkani-Hamed:2013jha,Cheung:2014dqa} as well as phenomenological applications of on-shell methods~\cite{Bern:2007dw,Berger:2008sj,Cheung:2015aba,Azatov:2016sqh}.
In particular,  the Lorentz covariant massive spinor formalism  was developed in Ref.~\cite{Arkani-Hamed:2017jhn}, which sets the stage for the application of on-shell constructive methods for scattering amplitudes with any masses or spins.  Many on-shell calculations for processes involving massive particles have been performed, including a bottom-up~\cite{Liu:2022alx} derivation of gauge invariance, and the enumeration of the  independent  contact terms in low-energy effective field theory~\cite{Durieux:2019eor,Balkin:2021dko,Liu:2023jbq,Dong:2022mcv}.


More recently a minor controversy has arisen since  ref. \cite{Christensen:2022nja} found that the constructive method residue did not seem to reproduce the correct amplitude for the simple tree-level process $e^- e^+\to\mu^- \mu^+$. We will resolve this issue here. After a brief review of the constructive method in section 2, we show in section \ref{sec:OFPT} how to use old fashioned perturbation theory (OFPT) to identify the physically relevant terms that must contribute to the residue at the on-shell pole. In section \ref{sec:4} show how this information can be used to eliminate irrelevant terms that arise in the constructive method, and which have been the source of previous confusion. We show that removing these irrelevant terms leads to the correct residue and hence the Feynman amplitude. We end with some brief conclusions and suggestions for future directions. An explanation of notation and conventions, as well as additional details regarding the constructive method, are relegated to two Appendices.

\section{Review of the Constructive Method}\label{sec:2}
First let us recall how on-shell, three-point, massless amplitudes can be glued to construct the residue of the four-point amplitude.
The on-shell massless three-point amplitudes, with fermions labeled by numbers and the spin-1 particle (to be glued) labeled by $V$ are:
\beq
\begin{aligned}
\mathcal{M} (1^{+\frac 12} 2^{- \frac 12} V^{+1})& = \frac{[1V]^2}{[12]}, \qquad \mathcal{M} (1^{ +\frac 12} 2^{- \frac 12} V^{-1}) = \frac{\langle 2V\rangle^2}{\langle12\rangle} \\
\mathcal{M} (3^{+\frac 12} 4^{- \frac 12} V^{+1}) &= \frac{[3 \tilde V]^2}{[34]}, \qquad \mathcal{M} (3^{ +\frac 12} 4^{- \frac 12}  V^{-1}) = \frac{\langle 4 \tilde V\rangle^2}{\langle34\rangle}~,
\end{aligned}
\eeq
with $p_V + p_1 + p_2 = 0$ and $p_{\tilde V} + p_3 + p_4 = 0$. Here we have adopted the all-momenta ingoing convention and the products of the  square (angle) brackets are defined using the standard spinor summation convention~\cite{Wess:1992cp,Dreiner:2008tw} (see Appendix~\ref{app:notation} for a summary of our notation and the conventions used in this paper):
\beq
[ij] = \tilde{\lambda}_{i,\dot \alpha} \tilde{\lambda}_j^{\dot \alpha}, \qquad \langle ij\rangle = \lambda^\alpha_i \lambda_{j,\alpha}~,
\eeq
 Momentum conservation tells us that $p_{\tilde V}  = - p_V$. When 3 and 4 are a different flavor from 1 and 2, the $s$-channel residue for the helicity amplitude $\mathcal (1^{+\frac 12} 2^{- \frac 12} 3^{+\frac 12} 4^{- \frac 12} )$ is obtained by gluing the three-point on-shell amplitude together:
\beq
\begin{aligned}
R_s (1^{+\frac 12} 2^{- \frac 12} 3^{+\frac 12} 4^{- \frac 12} ) &= \sum_{h = \pm 1} \mathcal{M} (1^{+\frac 12} 2^{- \frac 12} V^{h}) \times  \mathcal{M} (3^{ +\frac 12} 4^{- \frac 12} \tilde V^{-h})   \\
&= \frac{([1V] \langle4 \tilde V\rangle)^2}{[12]\langle 34\rangle} +  \frac{(\langle 2V \rangle [3 \tilde V])^2}{\langle 12\rangle [34]}  \\
& = \frac{([1 |p_V |4\rangle)^2}{[12]\langle 34\rangle} +  \frac{(\langle 2| p_V | 3])^2}{\langle 12\rangle [34]}  \\
& = -\frac{[1 |p_3 |4\rangle  [1 |p_2 |4\rangle}{[12]\langle 34\rangle} - \frac{\langle 2| p_1 | 3]  \langle 2| p_4 | 3]}{\langle 12\rangle [34]}  \\
&=  - 2  [13]\langle 24 \rangle~,
\end{aligned}
\eeq
where we have used the following analytical continuation:
\beq
|\tilde V] = |V], \qquad | \tilde V \rangle = - | V \rangle ~,
\eeq
and the identity:
\beq
[ i | p_j | k \rangle  =  [ i |_{\dot \alpha} p_j^{\dot \alpha \alpha} | k \rangle_\alpha= [ij]\langle  jk\rangle~.
\eeq
Now the helicity amplitude is simply
\beq
\mathcal{M} (1^{+\frac 12} 2^{- \frac 12} 3^{+\frac 12} 4^{- \frac 12} )  \propto\frac{R_s}{s} = - 2 \frac{ [13]\langle 24 \rangle}{s}~,
\eeq
which agrees with the Feynman diagram calculation.

Turning to massive fermions, one should be able to do a similar calculation using spinors that correspond to non-null momenta, as shown in  ref. \cite{Arkani-Hamed:2017jhn}, but the residue that ref. \cite{Christensen:2022nja} derived for $\mathcal{M}(e^-_1e^+_2\mu^-_3\mu^+_4)$ using the constructive method is:
\begin{equation}\label{eq:RNeil}
    R_s^\prime=\frac{1}{2m_e m_\mu}\Bigl[(u-t+2m_e^2+2m_\mu^2)[\mathbf{12}][\mathbf{34}]+2\bigl([\mathbf{12}][\mathbf{3}|p_2p_1|\mathbf{4}]+[\mathbf{1}|p_4p_3|\mathbf{2}][\mathbf{34}]\bigr)\Bigr],
\end{equation}
which does not  agree with the Feynman amplitude. We will see that this is because there are still (hidden) terms proportional to $s$ that need to be dropped in order to find the residue. 

\section{Scattering via OFPT}\label{sec:OFPT}

Before Feynman rules became the standard way of computing scattering amplitudes, OFPT was used. Dyson showed that the Feynman rules and OFPT are equivalent \cite{Dyson:1949bp}, so OFPT is now only used for pedagogic reasons. Unlike the Feynman rules, which use off-shell internal particles, intermediate states in OFPT are built with on-shell particles. The trade-off is that OFPT contains non-Lorentz invariant pieces at intermediate steps, and the interaction vertices do not conserve energy. However, Lorentz invariance is recovered after summing over all the pieces of the amplitude. Since the particles in OFPT are all on-shell, we expect looking at OFPT can give us some intuition for understanding the on-shell constructive method. In particular OFPT gives the unique on-shell analytic continuation of the Feynman amplitude, while the constructive method is only defined at isolated points in complex momentum space. The restriction to isolated points is essential for the residue method and also the reason why it is essential to remove terms in the lower-point amplitudes proportional to $s-m^2_{\rm pole}$.

In OFPT, we will take $p_3$ and $p_4$ to be outgoing. The $S$ matrix is obtained order by order using the interaction Hamiltonian \cite{Schwartz:2014sze} as a perturbation:
\begin{equation}
    \bra{f}S\ket{i}=\bra{f}H_\text{int}\ket{i}+\sum_n\frac{\bra{f}H_\text{int}\ket{n} \bra{n}H_\text{int}\ket{i}}{E_i-E_n}+ ...=\frac{1}{(2\pi)^3}\mathcal{M}_{fi}\delta^3(\V{p}_i-\V{p}_f)~.
\end{equation}
Graphically, the amplitude can be obtained by multiplying the vertices and summing over time-orderings of all diagrams, as we will show below.

For QED in Coulomb gauge, the interaction Hamiltonian has two terms, $H_\text{int}=H_T+H_\text{Coul}$, the first being the coupling to transverse photons and the second being the Coulomb term, which is necessary for obtain a Lorentz invariant amplitude. Following Weinberg's conventions \cite{Weinberg:1995mt},
\begin{equation}
    H_T=-\int d^3\V{x}\,\V{J}\cdot\V{A}\quad\text{and}\quad H_\text{Coul}=\frac{1}{2}\int d^3\V{x} d^3\V{y}\frac{J^0(\V{x})J^0(\V{y})}{4\pi|\V{x}-\V{y}|}~,
\end{equation}
where the bold symbols here represent three-vectors. The current $J^\mu=(J^0,\mathbf{J})$ expressed in terms of the Dirac field is:
\begin{equation}
    J^\mu = e\bar{\psi}\gamma^\mu\psi~,
\end{equation}
and the fields are:
\begin{equation}
    \begin{split}
        \psi(\V{x})=&(2\pi)^{-3/2}\int d^3\V{p}\sum_h[u_h(\V{p})b_h(\V{p})e^{-i\V{p}\cdot\V{x}}+v_h(\V{p})d^\dag_h(\V{p})e^{i\V{p}\cdot\V{x}}]\\
        A^\mu(\V{x})=&(2\pi)^{-3/2}\int \frac{d^3\V{p}}{\sqrt{2p^0}}\sum_{h=\pm}[\varepsilon^\mu_h(\V{p})a_h(\V{p})e^{-i\V{p}\cdot\V{x}}+\varepsilon^{\mu *}_h(\V{p})a^\dag_h(\V{p})e^{i\V{p}\cdot\V{x}}]~,
    \end{split}
\end{equation}
with normalization
\begin{equation}
    [a_h(\V{p}),a_{h'}^\dag(\V{p}')]=\{b_h(\V{p}),b_{h'}^\dag(\V{p}')\}=\{d_h(\V{p}),d_{h'}^\dag(\V{p}')\}=\delta_{hh'}\delta^3(\V{p}-\V{p}')~.
\end{equation}
 In this section we define the photon momentum to be $\V{k}$, which corresponds to $\V{p}_{\tilde{V}}$ in section \ref{sec:2}. Defining the on-shell photon energy by $\omega_{\V{k}}=\sqrt{\V{k}^2}$, the $\mathcal{O}(e^2)$ four-point vertex produced by  $H_\text{Coul}$ is:
\begin{equation}\label{eq:v_Coul}
    \includegraphics[height=2.5cm,valign=c]{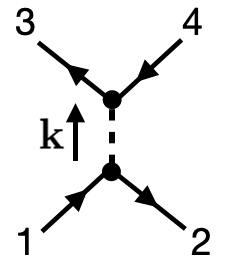}=\frac{e^2}{\omega_\V{k}^2}\bar{v}_4\gamma^0 u_3 \bar{v}_2 \gamma^0 u_1 =\frac{e^2}{\omega_\V{k}^2}\bigl(\braket{\mathbf{34}}+[\mathbf{34}]\bigr)\bigl(\braket{\mathbf{12}}+[\mathbf{12}]\bigr)~.
\end{equation}
For this vertex, $\V{k}$ is the sum of incoming fermion three-momenta, the non-covariant analog to the Feynman rule. The three-point vertex produced by $H_T$ is:
\begin{equation}\label{eq:v_T}
    \includegraphics[height=2cm,valign=c]{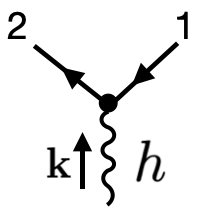}=\frac{e}{\sqrt{2\omega_\V{k}}}\bar{v}_2\slashed{\varepsilon}_h u_1\xrightarrow[]{s=0}\begin{cases}
        \displaystyle\frac{e}{\sqrt{\omega_{\V{k}}}\braket{kq_+}}\bigl(\braket{\mathbf{2}q_+}[k\mathbf{1}]+[\mathbf{2}k]\braket{q_+\mathbf{1}}\bigr),\quad h=+\\
        \\
        \displaystyle\frac{e}{\sqrt{\omega_{\V{k}}}[kq_-]}\bigl(\braket{\mathbf{2}k}[q_-\mathbf{1}]+[\mathbf{2}q_-]\braket{k\mathbf{1}}\bigr),\quad h=-
    \end{cases}~,
\end{equation}
up to crossing symmetry. This vertex is $\mathcal{O}(e)$ and (again) $\V{k}$ is the three-momentum of the on-shell photon. For $s=(p_1+p_2)^2=0$ we can  express the vertices in terms of  spinor helicity variables, by  using the formula in ref.~\cite{Liu:2022alx}, which uses the Weyl basis for $\gamma^\mu$. Here, $q_+$ and $q_-$ are arbitrary, null reference momenta. Note that we can only express $\slashed{\varepsilon}_h$ like this when $s=0$ and gauge invariance is, of course, preserved by the vertex.

For $\mathcal{M}\bigl(e^-(p_1)e^+(p_2)\rightarrow\mu^-(p_3)\mu^+(p_4)\bigr)$ the leading order amplitude is given by a Coulomb term plus terms with  transverse internal photons.
\begin{equation}
    \mathcal{M}=\includegraphics[height=2cm,valign=c]{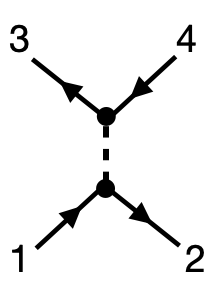}+\sum_{h=\pm}\Biggl(\includegraphics[height=2cm,valign=c]{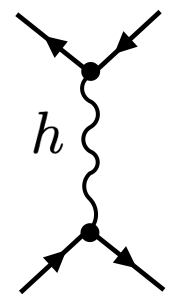}+\includegraphics[height=2cm,valign=c]{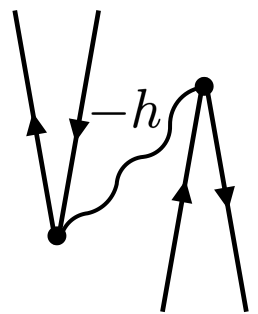}\Biggr)~.
\end{equation}
We will write these terms as $\mathcal{M}=\mathcal{M}_\text{Coul}+\mathcal{M}_+ +\mathcal{M}_-$ where $\mathcal{M}_\pm$ is the sum over time-orderings for two helicities, $\pm$, of the on-shell photon in the $\V{k}$ direction, $\V{k}=\V{p_1}+\V{p_2}$. Note that for the internal photon we combine time-orderings with opposite helicity because helicity is measured with respect to the photon momentum, so opposite helicity in opposite time-ordering corresponds to the same angular momentum.

Assembling the terms we we have:
\begin{equation}\label{3Ms_Coul}
    \begin{split}
        \mathcal{M}_\text{Coul}=&\frac{e^2}{\omega_\V{k}^2}(\braket{\mathbf{34}}+[\mathbf{34}])(\braket{\mathbf{12}}+[\mathbf{12}])\\
        \mathcal{M}_+=&\frac{2\omega_\V{k}}{s}\frac{e^2\bar{v}_2\slashed{\varepsilon}_+ u_1\bar{v}_4\slashed{\varepsilon}_+^* u_3}{\omega_\V{k}}\\
        \mathcal{M}_-=&\frac{2\omega_\V{k}}{s}\frac{e^2\bar{v}_2\slashed{\varepsilon}_- u_1\bar{v}_4\slashed{\varepsilon}_-^* u_3}{\omega_\V{k}}~,
    \end{split}
\end{equation}
where the $2\omega_\V{k}/s$ in the transverse terms comes from the $E_i-E_n$ in the denominator for second order transitions:
\begin{equation}
    \frac{1}{E_i-\omega_\V{k}}+\frac{1}{E_i-(2E_i+\omega_\V{k})}=\frac{-E_i-\omega_\V{k}+E_i-\omega_\V{k}}{(E_i-\omega_\V{k})(-E_i-\omega_\V{k})}=\frac{2\omega_\V{k}}{s}~.
\end{equation}
Note that in OFPT the energy of the photon $\omega_{\V{k}}\neq p_1^0+p_2^0=E_i$ in general because energy is not conserved at the vertices except at special values of complex momentum \cite{Badger_2005}. That is  if we choose specific complex external momenta such that $s=0$ then $\omega_{\V{k}}=E_i$. Another consequence of this is that some common formulae for $\varepsilon_h^\mu$ cannot be used. Under a Lorentz transformation, $\varepsilon_h^\mu$ transforms as $\varepsilon^\mu_h(p)\rightarrow\varepsilon^{\prime \mu}_h(p)+Cp^\mu$, that is with a term proportional to the photon's momentum \cite{Elvang:2013cua}. This term will not contribute to physical quantities due to the Ward-Takahashi identity. However, $\mathcal{M}_\pm$ are not amplitudes. They are just some pieces of the amplitude that are not Lorentz invariant by themselves. If we choose specific complex momenta such that $s=0$ and the photon's energy matches the initial energy $E_i$, we can express $\mathcal{M}\pm$ in terms of spinor helicity variables so that Lorentz invariance is manifest. The non-resonant Coulomb term, as we show below, is not relevant at the pole.

After pulling out the  $s$ pole, we choose specific momenta such that $s=0$. Eq.(\ref{3Ms_Coul}) then becomes:
\begin{equation}\label{eq:struc2look}
    \begin{split}
        s\mathcal{M}_\text{Coul}=&e^2 s \frac{(\braket{\mathbf{34}}+[\mathbf{34}])(\braket{\mathbf{12}}+[\mathbf{12}])}{\omega_\V{k}}\\
        s\mathcal{M}_+\bigg|_{s=0}=&e^2\frac{2(\braket{\mathbf{2}q_+}[k\mathbf{1}]+[\mathbf{2}k]\braket{q_+\mathbf{1}})(\braket{\mathbf{4}k}[q_-\mathbf{3}]+[\mathbf{4}q_-]\braket{k\mathbf{3}})}{\braket{kq_+}[kq_-]}\bigg|_{s=0}\\
        s\mathcal{M}_-\bigg|_{s=0}=&e^2\frac{2(\braket{\mathbf{2}k}[q_-'\mathbf{1}]+[\mathbf{2}q_-']\braket{k\mathbf{1}})(\braket{\mathbf{4}q_+'}[k\mathbf{3}]+[\mathbf{4}k]\braket{q_+'\mathbf{3}})}{\braket{kq_+'}[kq_-']}\bigg|_{s=0}
    \end{split}~.
\end{equation}
We see that after choosing particular complex momenta such that energy is conserved at the vertices ($s=0$), we find that $s\mathcal{M}_+$ and $s\mathcal{M}_-$ must saturate the residue while the Coulomb term does not contribute since $s\mathcal{M}_\text{Coul} \propto s =0$.

We can simplify $s\mathcal{M}_+$ as  follows: taking $q_+=q_-=q$ gives:
\begin{equation}\label{eq:Coul}
    \begin{split}
        &\frac{\braket{\mathbf{2}q}[k\mathbf{1}]+[\mathbf{2}k]\braket{q\mathbf{1}}}{\braket{kq}}\frac{\braket{\mathbf{4}k}[q\mathbf{3}]+[\mathbf{4}q]\braket{k\mathbf{3}}}{[kq]}\\
        =&\frac{\braket{\mathbf{2}q}[k\mathbf{1}]\braket{\mathbf{4}k}[q\mathbf{3}]+\braket{\mathbf{2}q}[k\mathbf{1}][\mathbf{4}q]\braket{k\mathbf{3}}+[\mathbf{2}k]\braket{q\mathbf{1}}\braket{\mathbf{4}k}[q\mathbf{3}]+[\mathbf{2}k]\braket{q\mathbf{1}}[\mathbf{4}q]\braket{k\mathbf{3}}}{\braket{kq}[kq]}\\
        =&\frac{\braket{\mathbf{2}q}[\mathbf{3}q][\mathbf{1}|k\ket{\mathbf{4}}+\braket{\mathbf{2}q}[\mathbf{4}q][\mathbf{1}|k\ket{\mathbf{3}}+\braket{q\mathbf{1}}[q\mathbf{3}][\mathbf{2}|k\ket{\mathbf{4}}+\braket{q\mathbf{1}}[q\mathbf{4}][\mathbf{2}|k\ket{\mathbf{3}}}{2q\cdot k}~.
    \end{split}
\end{equation}
We can apply the Schouten identity to the first two terms of Eq.(\ref{eq:Coul}) to get:
\begin{equation}
    \begin{split}
        \braket{\mathbf{2}q}[\mathbf{3}q][\mathbf{1}|k\ket{\mathbf{4}}=&-\braket{\mathbf{2}q}\bigl([q\mathbf{1}][\mathbf{3}|+[\mathbf{13}][q|\bigr)k\ket{\mathbf{4}}\\
        =&-[\mathbf{13}]\bra{\mathbf{2}}qk\ket{\mathbf{4}}-\braket{\mathbf{2}q}[q\mathbf{1}][\mathbf{3}|k\ket{\mathbf{4}}~,
    \end{split}
\end{equation}
\begin{equation}
\begin{split}
    \braket{\mathbf{2}q}[\mathbf{4}q][\mathbf{1}|k\ket{\mathbf{3}}=&-\braket{\mathbf{2}q}\bigl([q\mathbf{1}][\mathbf{4}|+[\mathbf{14}][q|\bigr)k\ket{\mathbf{3}}\\
    =&-[\mathbf{14}]\bra{\mathbf{2}}qk\ket{\mathbf{3}}-\braket{\mathbf{2}q}[q\mathbf{1}][\mathbf{4}|k\ket{\mathbf{3}}~,
\end{split}
\end{equation}
where their second terms cancel:
\begin{equation}
    [\mathbf{3}|k\ket{\mathbf{4}}+[\mathbf{4}|k\ket{\mathbf{3}}=m_\mu\bigl(\braket{\mathbf{43}}-[\mathbf{43}]+\braket{\mathbf{34}}-[\mathbf{34}]\bigr)=0~.
\end{equation}
A similar cancellation happens to the last two terms in Eq.(\ref{eq:Coul}). Therefore,  $s\mathcal{M}_+$  is given by:
\begin{equation}
s\mathcal{M}_+=-e^2\frac{[\mathbf{13}]\bra{\mathbf{2}}qk\ket{\mathbf{4}}+[\mathbf{14}]\bra{\mathbf{2}}qk\ket{\mathbf{3}}+[\mathbf{23}]\bra{\mathbf{1}}qk\ket{\mathbf{4}}+[\mathbf{24}]\bra{\mathbf{1}}qk\ket{\mathbf{3}}}{2q\cdot k}~.
\end{equation}
Similarly, $s\mathcal{M}_-$ becomes:
\begin{equation}
    s\mathcal{M}_-=-e^2\frac{[\mathbf{13}]\bra{\mathbf{2}}kq\ket{\mathbf{4}}+[\mathbf{14}]\bra{\mathbf{2}}kq\ket{\mathbf{3}}+[\mathbf{23}]\bra{\mathbf{1}}kq\ket{\mathbf{4}}+[\mathbf{24}]\bra{\mathbf{1}}kq\ket{\mathbf{3}}}{2q\cdot k}~.
\end{equation}
Adding them together and using $qk+kq=(2q\cdot k) \mathds{1}$, we see the $2q\cdot k$ in the denominator is cancelled. Thus, the residue becomes:
\begin{equation}\label{eq:Mresidue}
    s\mathcal{M}=-e^2\bigl([\mathbf{13}]\braket{\mathbf{24}}+[\mathbf{14}]\braket{\mathbf{23}}+[\mathbf{23}]\braket{\mathbf{14}}+[\mathbf{24}]\braket{\mathbf{13}}\bigr)~,
\end{equation}
which agrees with the Feynman amplitude. The extra minus sign comes from $\ket{-p}=-\ket{p}$ when we take $p_3$ and $p_4$ to be outgoing. Thus we know the structure to look for in the constructive method calculation.

\begin{figure}[htb]
\includegraphics[width=12cm]{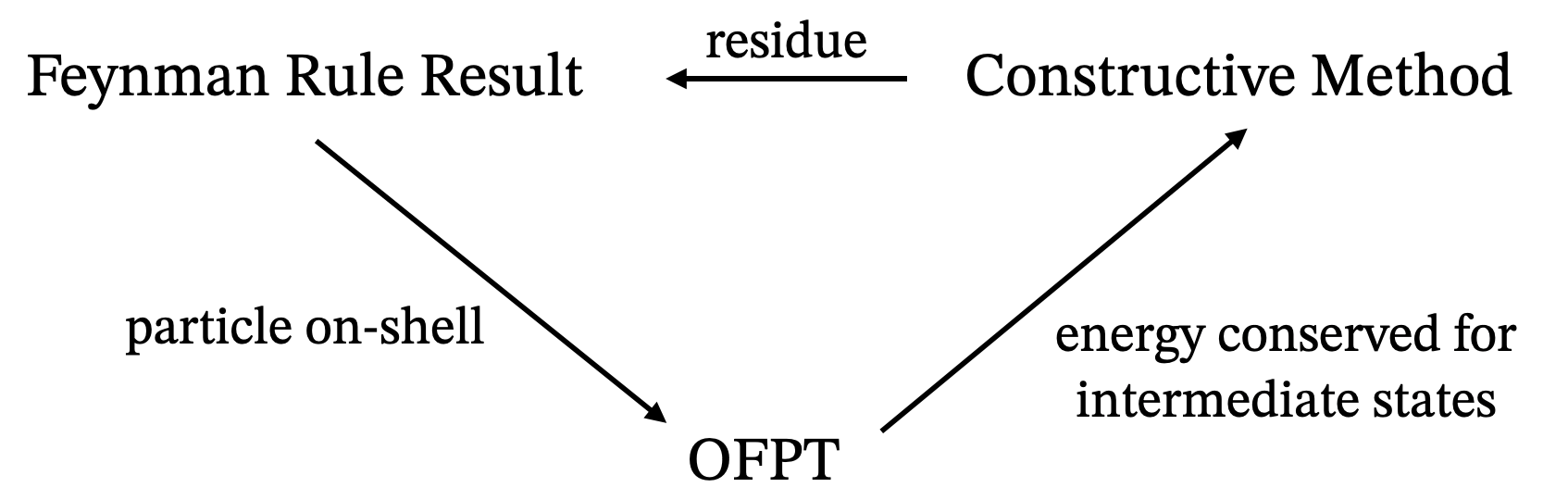}
\centering
\centering
\caption{Relation between Feynman amplitude, the OFPT amplitude, and the constructive method amplitude.}
\label{box}
\end{figure}

In Fig. \ref{box} we display the relation between the three approaches to finding the amplitude. Since OFPT provides a unique on-shell analytic continuation of the Feynman amplitude, if we restrict to a particular point in complex momentum space where energy is conserved at the vertices then it must reproduce exactly the results of the constructive method, term by term, since we have isolated the contribution of individual on-shell photons with definite helicity. Thus we do not even have to calculate the Coulomb term, we can still obtain the Feynman amplitude by finding the residue, if we consistently drop the terms higher order in $s$.

\section{Constructive Method for Massive Fermions}\label{sec:4}

The on-shell massive three-point amplitudes for the particles $(1,2,k)$ with massive fermions and a photon are given by:
\beq
\mathcal{M}(1^{I_1} 2^{I_2} k^{+1}) = x_{12} \braket{\mathbf{12}}, \qquad \mathcal{M}(1^{I_1} 2^{I_2} k^{-1}) = \tilde {x}_{12} [\mathbf{12}]~,
\eeq
where we have used the bold notation introduced in Ref.~\cite{Arkani-Hamed:2017jhn}. Similar formulae hold for particle $(3,4, k)$. The $x$-factor is defined as: 
\begin{equation}
    x_{12}=\frac{\bra{q}p_2|(-k)]}{m_e\braket{q (-k)}}, \qquad         \Tilde{x}_{12}= \frac{[\Tilde{q}|p_2\ket{(-k)}}{m_e[\Tilde{q} (-k)]}~,
\end{equation}
and similar definitions apply to the (3,4,$k$):
\beq\label{xdefinition}
    x_{34}=\frac{\bra{q}p_4|k]}{m_\mu\braket{q k}}
    \quad
    \Tilde{x}_{34}=\frac{[\Tilde{q}|p_4\ket{k}}{m_\mu[\Tilde{q}k]}~,
    \eeq
 where in order to easily compare with the OFPT calculation, we have defined $k = p_1 + p_2 = - p_3 - p_4$ and $q_{\alpha}, \tilde{q}_{\dot \alpha} $ are reference spinors.  One can show explicitly that the $x$-factors are independent of the choices of the reference spinors for the on-shell momenta (see Appendix~\ref{app:xdef} for explicit demonstration). Note that here we have chosen the same reference spinors in the definitions of $x_{12} (\tilde{x}_{12})$ and  $x_{34} (\tilde{x}_{34})$ for convenience.
 Before we move on to calculate the four-point amplitude, let us take a closer look at the on-shell, massive three-point amplitude. Note that by using the Schouten identity, we can rewrite the amplitude $\mathcal{M}(1^{I_1} 2^{I_2} k^{+1})$ as follows:
\begin{equation}
\label{eq:x12schouten}
\begin{split}
    x_{12}\braket{\mathbf{12}}=&\frac{\bra{q}p_2|(-k)]}{m_e\braket{q(-k)}}\braket{\mathbf{12}}\\
    =&\bigl(\braket{\mathbf{2}q}\bra{\mathbf{1}}+\braket{q\mathbf{1}}\bra{\mathbf{2}}\bigr)\frac{p_2|k]}{m_e\braket{qk}}\\
    =&\frac{\braket{\mathbf{2}q}[\mathbf{1}k]-\braket{q\mathbf{1}}[\mathbf{2}k]}{\braket{qk}} + \frac{\braket{\mathbf{2}q}\bra{\mathbf{1}}k|k]}{m_e\braket{qk}}~,
\end{split}
\end{equation}
where we used $k=p_1+p_2$ in the last step and $\bra{\mathbf{i}} p_i = -m_i [\mathbf{i}|$. The second term can be written as:
\begin{equation}
    \frac{\braket{\mathbf{2}q}\bra{\mathbf{1}}k|k]}{m_e\braket{qk}}=\frac{\braket{\mathbf{2}q}\bra{\mathbf{1}}k|k]\braket{kq}}{m_e\braket{qk}\braket{kq}}=s\frac{\braket{q\mathbf{2}}\braket{\mathbf{1}q}}{m_e\braket{qk}^2}~,
\end{equation}
which is proportional to $s$.
 Note that the second term in the last line of  Eq.~(\ref{eq:x12schouten}) vanishes for an on-shell photon ($s=0$), and can cause some confusion when calculating the four-point scattering amplitude. The high energy limit of the on-shell  massive three-point amplitude can be obtained by focusing on the first term alone. For example, the helicity amplitude $\mathcal{M}(1^{+\frac 12} 2^{-\frac 12} k^{+1}) $ can be obtained as follows:
  \begin{equation}
 \label{eq:x12he}
    \begin{split}
\mathcal{M}(1^{+\frac 12} 2^{-\frac 12} k^{+1}) &=
   \frac{\braket{2q_{12}}[1(-k)]-\braket{q_{12}\eta_1}[\tilde{\eta}_2 (-k)]}{\braket{q_{12}(-k)}}  \rightarrow     \frac{\braket{21}[1(-k)]}{\braket{1(-k)}} =   \frac{[1(-k)]^2}{ [12]}   ~.
    \end{split}
\end{equation}
 We can perform similar operations to $\Tilde{x}_{34}[\mathbf{34}]$ and the other combination of $x$-factors multiplying spinor overlaps. At the end of the day, we have:
\begin{equation}
\label{eq:xfactorsnew}
    \begin{split}
        x_{12}\braket{\mathbf{12}}=&\frac{\braket{\mathbf{2}q}[\mathbf{1}k]-\braket{q\mathbf{1}}[\mathbf{2}k]}{\braket{qk}}+s\frac{\braket{q\mathbf{2}}\braket{\mathbf{1}q}}{m_e\braket{qk}^2}\\
        \Tilde{x}_{12}[\mathbf{12}]=&\frac{[\mathbf{2}q]\braket{\mathbf{1}k}-[q\mathbf{1}]\braket{\mathbf{2}k}}{[qk]}+s\frac{[q\mathbf{2}][\mathbf{1}q]}{m_e[qk]^2}\\
        x_{34}\braket{\mathbf{34}}=&\frac{\braket{q\mathbf{4}}[\mathbf{3}k]-\braket{\mathbf{3}q}[\mathbf{4}k]}{\braket{qk}}+s\frac{\braket{\mathbf{4}q}\braket{\mathbf{3}q}}{m_\mu\braket{qk}^2}\\
        \Tilde{x}_{34}[\mathbf{34}]=&\frac{[q\mathbf{4}]\braket{\mathbf{3}k}-[\mathbf{3}q]\braket{\mathbf{4}k}}{[qk]}+s\frac{[\mathbf{4}q][\mathbf{3}q]}{m_\mu[qk]^2}~.
    \end{split}
\end{equation}
So we see that the  $x$-factor multiplied by the corresponding spinor overlap gives the OFPT vertex plus an  $\mathcal{O}(s)$ term. Using the constructive method, the residue at the $s$-pole of $\mathcal{M}(e^-e^+\rightarrow \mu^-\mu^+)$ with all particles in-coming is \cite{Christensen:2022nja}:
\begin{equation}\label{eq:Rs1}
    R_s(e^-_1e^+_2\mu^-_3\mu^+_4)=x_{12}\Tilde{x}_{34}\braket{\mathbf{12}}[\mathbf{34}]+\Tilde{x}_{12}x_{34}[\mathbf{12}]\braket{\mathbf{34}}~.
\end{equation}
Note that according to  Eq.~(\ref{eq:xfactorsnew}), there will be terms in the above formula with powers of $s$ between 0 and 2, while the residue corresponds to the term that survives at $s=0$. 

By only keeping the $\mathcal{O}(s^0)$ term,  the first term in Eq.(\ref{eq:Rs1}) becomes:
\begin{equation}\label{eq:Rs3}
\begin{split}
    &\frac{\braket{\mathbf{2}q}[\mathbf{1}k]-\braket{q\mathbf{1}}[\mathbf{2}k]}{\braket{qk}}\frac{[q\mathbf{4}]\braket{\mathbf{3}k}-[\mathbf{3}q]\braket{\mathbf{4}k}}{[qk]}\\
    =&\frac{(\braket{\mathbf{2}q}[k\mathbf{1}]+[\mathbf{2}k]\braket{q\mathbf{1}})(\braket{\mathbf{4}k}[q\mathbf{3}]+[\mathbf{4}q]\braket{k\mathbf{3}})}{\braket{kq}[kq]}~,
\end{split}
\end{equation}
which agrees with the OFPT result in Eq.(\ref{eq:struc2look}), corresponding to the gluing of vertices in Eq.(\ref{eq:v_T}). Agreement is also found for the second term in Eq.(\ref{eq:Rs1}). We have shown that the sum gives Eq.(\ref{eq:Mresidue}). Thus, the residue is:
\begin{equation}\label{Rs4}
    R_s=[\mathbf{14}]\braket{\mathbf{23}}+[\mathbf{13}]\braket{\mathbf{24}}+\braket{\mathbf{24}}[\mathbf{13}]+\braket{\mathbf{23}}[\mathbf{14}]~,
\end{equation}
there is no minus sign because we are taking all momenta to be incoming here. After putting back the $e^2$ coupling and $s$-pole, this gives us the correct Feynman amplitude.

It can be shown that Eq.(\ref{eq:RNeil}) satisfies
\begin{equation}
    R_s^\prime=R_s+\frac{s}{2m_em_\mu}\bigl([\mathbf{12}][\mathbf{34}]-2[\mathbf{14}][\mathbf{23}]\bigr)~,
\end{equation}
so that it agrees with the Feynman amplitude after dropping the term proportional to $s$.

For the sake of completeness we  conclude this subsection by providing the $\mathcal{O}(s)$ and $\mathcal{O}(s^2)$ terms in Eq.(\ref{eq:Rs1}):
\begin{equation}
    \begin{split}
        \mathcal{O}(s)=s\biggl(&\frac{\braket{\mathbf{2}q}[\mathbf{1}k]-\braket{q\mathbf{1}}[\mathbf{2}k]}{\braket{qk}}\frac{[\mathbf{4}q][\mathbf{3}q]}{m_\mu[qk]^2}+\frac{[q\mathbf{4}]\braket{\mathbf{3}k}-[\mathbf{3}q]\braket{\mathbf{4}k}}{[qk]}\frac{\braket{q\mathbf{2}}\braket{\mathbf{1}q}}{m_e\braket{qk}^2}\\
        &+\frac{[\mathbf{2}q]\braket{\mathbf{1}k}-[q\mathbf{1}]\braket{k\mathbf{2}}}{[qk]}\frac{\braket{\mathbf{4}q}\braket{\mathbf{3}q}}{m_\mu\braket{qk}^2}+\frac{\braket{q\mathbf{4}}[\mathbf{3}k]-\braket{\mathbf{3}q}[\mathbf{4}k]}{\braket{qk}}\frac{[q\mathbf{2}][\mathbf{1}q]}{m_e[qk]^2}\biggr)~,
    \end{split}
\end{equation}
and
\begin{equation}
    \mathcal{O}(s^2)=s^2\biggl(\frac{\braket{q\mathbf{2}}\braket{\mathbf{1}q}[\mathbf{4}q][\mathbf{3}q]+[q\mathbf{2}][\mathbf{1}q]\braket{\mathbf{4}q}\braket{\mathbf{3}q}}{4m_e m_\mu (q\cdot k)^2}\biggr)~.
\end{equation}

\section{Conclusions\label{s.con}}

We have demonstrated that after expanding the glued three-point amplitudes in powers of $s-m^2_{\rm pole}$ (for the photon $m_{\rm pole}=0$) we can find residue given by the $\mathcal{O}((s-m^2_{\rm pole})^0)$ term, and that this result agrees with the Feynman amplitude, as it should.
We have also seen that OFPT provides an on-shell analytic continuation that reproduces the constructive method term by term, thus extending the on-shell amplitude to all complex momentum, not just isolated kinematic points in complex momentum space. We have seen that in this example the constructive method corresponds precisely to the contribution of the transverse on-shell photons of OFPT at this special point. 

It would be interesting to see if OFPT can be useful in extending recursive methods to amplitudes with massive particles.

\section*{Acknowledgments}

We thank Neil Christensen, Markus Luty and Matthew Schwartz for helpful discussions.
This work was supported in part by the DOE under grant DE-SC-000999 and the work of DL was also supported partially by  the DOE under grant No.~DE-SC0007914 and in part by PITT PACC. Part of the work was performed at Aspen Center for Physics, which is supported by NSF grant PHY-2210452.


\appendix
\numberwithin{equation}{section}
\renewcommand{\theequation}{\thesection.\arabic{equation}}

\section{Notations and Conventions}
\label{app:notation}

In this appendix, we collect the notations and conventions used throughout the paper. We will use the mostly minus metric $\eta_{\mu\nu} = \text{diag}(1,-1,-1,-1)$. Repeated indices are summed over unless otherwise mentioned. The $2\times 2$ matrices corresponding to a four-momentum are defined as:
\begin{equation}
    p_{\alpha\dot{\alpha}}=p_\mu\sigma^\mu_{\alpha\dot{\alpha}}\quad\text{and}\quad p^{\alpha\dot{\alpha}}=p^\mu\bar{\sigma}_\mu^{\dot{\alpha}\alpha}~,
\end{equation}
where the $\sigma^\mu$ matrices are defined as:
\begin{equation}
    \sigma^\mu=(\mathds{1},\boldsymbol{\sigma})\quad\text{and}\quad\bar{\sigma}^\mu=(\mathds{1},-\boldsymbol{\sigma})~,
\end{equation}
with $\boldsymbol{\sigma}=(\sigma_x,\sigma_y,\sigma_z)$ being the Pauli matrices. The identity
\begin{equation}
    \sigma^\mu_{\alpha\dot{\alpha}}\bar{\sigma}^{\nu\dot{\alpha}\beta}+\bar{\sigma}^\nu_{\alpha\dot{\alpha}}\sigma^{\mu\dot{\alpha}\beta}=2\eta^{\mu\nu}\delta_\alpha^\beta~,
\end{equation}
allows us to write
\begin{equation}
    p_{\alpha\dot{\alpha}}k^{\dot{\alpha}\beta}+k_{\alpha\dot{\alpha}}p^{\dot{\alpha}\beta}=2(p\cdot k)\delta_\alpha^\beta~.
\end{equation}
We will often suppress the spinor indices and write equations like this  as $pk+kp=2(p\cdot k)$ instead.

\subsection{Spinor-Helicity Variables for Massless Particle}

For a massless particle, $p^2=0$ and we can factorize $p_{\alpha\dot{\alpha}}$ into a product of two spinors.
\begin{equation}
    p_{\alpha\dot{\alpha}}=\ket{p}_\alpha [p|_{\dot{\alpha}}~,
\end{equation}
which, suppressing the spinor indices, is $p=\ket{p}[p|$. Square or angle brackets indicates the chirality of the spinor. We can raise or lower the spinor index using the fully anti-symmetric $2\times 2$ tensor $\varepsilon^{\alpha\beta}$.
\begin{equation}
    \bra{p}^\alpha=\varepsilon^{\alpha\beta}\ket{p}_\beta\quad\text{and}\quad |p]^{\dot{\alpha}}=\varepsilon^{\dot{\alpha}\dot{\beta}}[p|_{\dot{\beta}}~,
\end{equation}
where
\begin{equation}
    \varepsilon^{\alpha\beta}\varepsilon_{\beta\gamma}=\delta^\alpha_\gamma~.
\end{equation}
In this paper, we take $\varepsilon^{12}=-\varepsilon_{12}=1$. We use the North-South  and South-North conventions for undotted and dotted spin contractions  \cite{Dreiner:2008tw}. The contraction between two spinors is thus:
\begin{equation}
    \varepsilon^{\alpha\beta}\ket{p}_\alpha\ket{k}_\beta=\bra{k}^\alpha\ket{p}_\alpha=\braket{kp}
    \quad\text{and}\quad
    \varepsilon_{\dot{\alpha}\dot{\beta}}|p]^{\dot{\alpha}}|k]^{\dot{\beta}}=[k|_{\dot{\alpha}}|p]^{\dot{\alpha}}=[kp]~.
\end{equation}
Note that due to $\varepsilon^{\alpha\beta}$ being anti-symmetric, $\braket{kp}=-\braket{pk}$ and $[kp]=-[pk]$. We can also have contractions between $p_{\alpha\dot{\alpha}}$ matrices and spinors.
\begin{equation}
    [q|_{\dot{\beta}}p^{\dot{\beta}\alpha}\ket{k}_{\alpha}=[q|p\ket{k}~.
\end{equation}

For a four momentum $p^\mu=(E,E\sin{\theta}\cos{\phi},E\sin{\theta}\sin{\phi},E\cos{\theta})$, we can choose the spinors to be  \cite{Dreiner:2008tw}:
\begin{equation}
    \ket{p}=\sqrt{2E}\begin{bmatrix}
        -s^*\\
        c
    \end{bmatrix}
    \quad
    |p]=\sqrt{2E}\begin{bmatrix}
        -s\\
        c^*
    \end{bmatrix}~,
\end{equation}
where
\begin{equation}
    s\equiv\sin{\frac{\theta}{2}}e^{i\phi} \quad
    \text{and}\quad
    c\equiv \cos{\frac{\theta}{2}} e^{i\phi}~.
\end{equation}
This leads to the analytic continuation
\begin{equation}
    \ket{(-p)}=-\ket{p}\quad|(-p)]=|p]
\end{equation}
under $E\rightarrow-E$, $\theta\rightarrow\pi-\theta$ and $\phi\rightarrow\phi+\pi$.

\subsection{Spinor-Helicity Variables for Massive Particle}

For particle with mass $m$, $p^2=m^2$ and we cannot factorize $p_{\alpha\dot{\alpha}}$ into two spinors. Instead, we can write it as the sum of two such pairs.
\begin{equation}
    p_{\alpha\dot{\alpha}}=\ket{p^I}[p_I|=\ket{p^+}[p_+|+\ket{p^-}[p_-|~,
\end{equation}
where $I=+,-$ is the little group index and can also be raised/lowered by $\varepsilon^{IJ}$. We will drop the momentum symbol in the spinors for simplicity.
\begin{equation}
    \ket{p_i^I}=\ket{i^I}~.
\end{equation}
In this notation we have:
\begin{equation}
    \ket{i^I}_\alpha\ket{i_I}_\beta=m_i\varepsilon_{\alpha\beta},
    \quad
    [i^I|_{\dot{\alpha}}[i_I|_{\dot{\beta}}=m\varepsilon_{\dot{\alpha}\dot{\beta}},
    \quad
    \braket{i^I i^J}=-m_i\varepsilon^{IJ}\quad
    \text{and}
    \quad
    [i_L i_K]=-m_i\varepsilon_{LK}~.
\end{equation}
The Dirac equation then becomes:
\begin{equation}
    \bra{i^I}p_i=-m_i[i^I|\quad\text{and}\quad p_i|i^I]=m_i\ket{i^I}~.
\end{equation}
The fully anti-symmetric tensor $\varepsilon_{\alpha\beta}$ also satisfies the Schouten identity \cite{Elvang:2013cua,Christensen:2019mch}:
 \begin{equation}
     \varepsilon_{\alpha\beta}\varepsilon_{\gamma\delta}=\varepsilon_{\alpha\gamma}\varepsilon_{\beta\delta}-\varepsilon_{\alpha\delta}\varepsilon_{\beta\gamma}~.
 \end{equation}
 After contracting with spinors, the Schouten identity becomes:
 \begin{equation}
     \ket{i}\braket{jk}+\ket{j}\braket{ki}+\ket{k}\braket{ij}=0~,
 \end{equation}
 which works the same for massless, massive, right or left handed spinors.

\section{Definition of the $x$-factor}
\label{app:xdef}

In this Appendix we show that changing the reference spinor in the $x$-factor does not generate extra terms proportional to $s$. Starting with the definition in Eq. (\ref{xdefinition}) and applying the Schouten identity gives:
\begin{equation}
    \begin{split}
        x_{34}=&\frac{\bra{q_{34}}p_4|k]}{m_\mu\braket{q_{34}k}}=\frac{\bra{q_{34}}p_4|k]}{m_\mu\braket{q_{34}k}}\frac{\braket{q'k}}{\braket{q'k}}\\
        =&-(\braket{kq_{34}}\bra{q'}+\braket{q_{34}q'}\bra{k})\frac{p_4|k]}{m_\mu\braket{q_{34}k}\braket{q'k}}\\
        =&\frac{\bra{q'}p_4|k]}{m_\mu\braket{q'k}}+s\frac{\braket{q_{34}q'}}{m_\mu\braket{q_{34}k}\braket{q'k}}~,
    \end{split}
\end{equation}
where $\bra{k}p_4|k]=2p_4\cdot k = -s$. It seems like changing the reference spinor will generate a $\mathcal{O}(s)$ term, but we can use $p_4=-k-p_3$ to rewrite the first term as:
\begin{equation}
    \begin{split}
        x_{34}=&\frac{\bra{q'}(-k-p_3)|k]}{m_\mu\braket{q'k}}+s\frac{\braket{q_{34}q'}}{m_\mu\braket{q_{34}k}\braket{q'k}}\\
        =&\frac{\bra{q'}(-p_3)|k]}{m_\mu\braket{q'k}}-\frac{\bra{q'}k|k]\braket{kq_{34}}}{m_\mu\braket{q'k}\braket{kq_{34}}}+s\frac{\braket{q'q_{34}}}{m_\mu\braket{q'k}\braket{kq_{34}}}\\
        =&\frac{\bra{q'}(-p_3)|k]}{m_\mu\braket{q'k}}-s\frac{\braket{q'q_{34}}}{m_\mu\braket{q'k}\braket{kq_{34}}}+s\frac{\braket{q'q_{34}}}{m_\mu\braket{q'k}\braket{kq_{34}}}=\frac{\bra{q'}(-p_3)|k]}{m_\mu\braket{q'k}}~,
    \end{split}
\end{equation}
where we used $k^2=s$. We see that we can eliminate the $\mathcal{O}(s)$ term by writing:
\begin{equation}
    x_{34}=\frac{\bra{q_{34}}p_4|k]}{m_\mu\braket{q_{34}k}}=\frac{\bra{q'}(-p_3)|k]}{m_\mu\braket{q'k}}
\end{equation}
Similarly, we can use
\begin{equation}
    \frac{\bra{q'}(-p_3)|k]}{m_\mu\braket{q'k}}=\frac{\bra{q''}p_4|k]}{m_\mu\braket{q''k}}
\end{equation}
without generating an $\mathcal{O}(s)$ term. Therefore, we can do both of these in a row and get:
\begin{equation}
    x_{34}=\frac{\bra{q_{34}}p_4|k]}{m_\mu\braket{q_{34}k}}=\frac{\bra{q'}(-p_3)|k]}{m_\mu\braket{q'k}}=\frac{\bra{q''}p_4|k]}{m_\mu\braket{q''k}}~.
\end{equation}
Thus changing the reference spinor does not generate an extra $\mathcal{O}(s)$ term.


\end{document}